# A DFT study of B-doped graphene as a metal-anchor: effects of oxidation and strain

Nikola Veličković[1], Natalia V. Skorodumova[2], Ana S. Dobrota[1]*

[1] University of Belgrade - Faculty of Physical Chemistry, Studentski trg 12-16, 11158 Belgrade, Serbia

[2] Applied Physics, Division of Materials Science, Department of Engineering Sciences and Mathematics, Luleå University of Technology, 97187 Luleå, Sweden

*Corresponding author:

Dr. Ana S. Dobrota, Assistant Professor

University of Belgrade – Faculty of Physical Chemistry

Studentski trg 12-16, 11158 Belgrade, Serbia

Email: ana.dobrota@ffh.bg.ac.rs

**Abstract**

In this work, we present a systematic DFT investigation of the interaction between B-doped graphene and four selected metals: Mg and Zn, relevant for next-generation metal-ion batteries, and Cu and Pt, important for single-atom catalysis. Three different boron doping concentrations were considered to elucidate how dopant density influences the binding strength, charge transfer, and electronic structure of the resulting systems. In addition, the effects of biaxial strain and surface oxidation were examined to assess their impact on the reactivity and stability of B-doped graphene. The results show that boron doping substantially enhances graphene's affinity toward metal adsorption, though the extent and nature of this effect depend strongly on the metal type and doping level. For some of the metals investigated, the interaction is found to be almost entirely charge-transfer driven, with minimal orbital hybridization. Mechanical strain is found to enable fine-tuning of the metal/substrate interaction, while surface oxidation introduces a more pronounced effect by enabling direct interaction between metal atoms and oxygen functional groups in most cases, thereby significantly altering adsorption geometry and strength. These findings provide valuable insights into the design of boron-doped graphene materials for energy conversion and storage applications.

# 1. Introduction

Graphene has attracted extensive attention due to its exceptional electrical, mechanical, and thermal properties, as well as its high surface area and chemical stability.[1] However, its intrinsic chemical inertness limits its application in catalysis and adsorption-based processes. Heteroatom doping and strain engineering are well-known strategies for altering the reactivity of surfaces. Dopant atoms change the local electronic structure, induce charge redistribution, and generate sites with altered binding affinity compared to defect-free (pristine) graphene.[2,3] Boron (B) doping is especially interesting for several reasons. First, boron has an atomic radius similar to carbon but is electron-deficient (p-type dopant), which produces localized positive charge centers and modifies π-electron distribution in the graphene plane. [2,4,5] These features increase the chemical affinity toward electron-rich adsorbates and metal adatoms. Therefore, holes induced by B-doping might help stabilize isolated metal atoms via stronger metal-support interactions, thus inhibiting aggregation. Second, B-doped graphene retains the desirable conductive framework of graphene while introducing tunable reactivity by varying dopant concentration and local bonding environment. These attributes make B-doped graphene a promising scaffold for anchoring single metal atoms or small clusters with improved stability and tailored electronic properties. For example, B doping was found to improve graphene's interaction with transition metals Sc, Ti, and V.[6] Next, mechanical strain is known to be another powerful parameter in tuning the electronic structure and catalytic performance of materials, including graphene.[7] Even small deformations can alter the bond lengths, orbital overlaps, and charge distribution within the lattice, thereby influencing the adsorption strength and reaction pathways of surface species. Since realistic operating conditions, such as substrate mismatch, thermal expansion, or mechanical manipulation, often introduce tensile or compressive strain, understanding its effect becomes essential for reliable design of materials suitable for chosen applications. Mechanical strain was shown to enable functionalization of graphene for both p- and n-type dopants, while unstrained graphene showed negligible reactivity towards them.[7] Uniaxial strain above 20% was found to open a gap in graphene, while biaxial strain preserved its crystal symmetry, did not open a gap, but changed the slope of the Dirac cones.[8] The tensile strain of about 10% applied in graphene was shown to greatly increase the adsorption energies of various kinds of metal clusters.[9] Tensile strain was also found to enhance the binding strength of sodium onto pristine, N- and P-doped graphene.[10]

In this contribution, we investigate the interactions of four chosen types of metal atoms with B-doped graphene: Mg, Cu, Zn and Pt. These metals were chosen to sample a range of chemical behavior: from alkali-earth (Mg) to late-transition (Cu, Zn) and noble (Pt), enabling identification of trends in binding strength, charge redistribution, and potential catalytic relevance. We chose Mg and Zn as the representatives of metals promising for novel metal-ion batteries. Both have higher volumetric energy density compared to widely used lithium because they carry a +2 charge per ion, allowing for more charge transfer per ion.[11] However, their ions are larger compared to $Li^+$, which makes it harder for them to move through electrodes and electrolytes. Magnesium is more abundant than lithium and is widely available in the Earth's crust and seawater. This makes it cheaper and more sustainable. Zn is also more accessible and relatively cheap resource, also known for its non-toxicity and chemical stability in aqueous media.[12] Riyaz *et al.* theoretically predicted B-doped graphene as an efficient anode material for magnesium-ion battery, with good electrical and ionic conductivity and high cycling stability.[13] Cu and Pt are chosen here as interesting candidates for single atom catalysts (SACs).[14] In traditional metal catalysts, bulk metals or metal nanoparticles, many metal atoms are buried inside and don't participate in the reaction. In SACs each metal atom is fully exposed, maximizing its catalytic efficiency. This significantly reduces the amount of metal needed, which is especially important for SACs containing expensive noble metals. Platinum is scarce and expensive. Using single atoms means higher catalytic performance per Pt atom, reducing overall usage and price. This makes SACs more economically viable for industrial applications. Pt-based catalysts are generally considered to be the most effective electrocatalysts for the Hydrogen Evolution Reaction (HER).[15] B-doped graphene decorated with Pt nanoparticles has been shown to have good activity for Oxygen Reduction Reaction (ORR).[16] Copper is much cheaper than noble metals and is the catalyst of choice for $CO_2$ reduction ($CO_2R$), since it is the only elemental metal that can further reduce $CO_2$ beyond CO or formate to form multi-carbon ($C_2^+$) products, such as ethylene, ethanol, acetate, propanol, methane, which are valuable chemicals and energy-dense fuels. This is due to copper's optimal (intermediate) strength of CO binding. CO binding strength was proposed as a descriptor for $CO_2R$ activity due to its linear correlation with reaction barriers of the $CO_{ads}$ reduction step on multiple metallic surfaces.[17] Therefore, the SACs containing Cu are an important possibility to investigate. Recently, graphene with a $ZnB_4$ defect was proposed as one of the hypothetical SACs that might be efficient electrocatalysts for $CO_2$ reduction reaction.[18] Additionally, enhanced metal binding to graphene can be interesting from the viewpoint of removing copper and zinc, harmful for the environment, from industrial wastewater.[19]

Despite all the mentioned progress, a systematic, comparative understanding of how different metal atoms (spanning main-group Mg, late transition metals Cu and Zn, and noble Pt) interact with B-doped graphene is still needed to guide rational catalyst design. Moreover, such systematization could reveal which metal / B-doped graphene combinations would be likely to remain atomically dispersed under working conditions, provide appropriate binding strengths for targeted adsorbates or reaction intermediates, and tune active-site electronic structure toward desired reaction energetics. We employ Density Functional Theory (DFT) calculations to study adsorption geometries, adsorption energies, charge transfer, and electronic structure of Mg, Cu, Zn and Pt atoms on boron-doped graphene. Our results clarify how boron concentration, strain, and surface oxidation modify metal-graphene interactions and provide guidance for selecting promising doped graphene materials for applications in electrocatalysis and rechargeable metal-ion devices.

## 2. Computational details

Density Functional Theory (DFT) calculations were conducted using the open-source suite Quantum ESSPRESO.[20,21] As we were interested in the interaction of chosen metals with graphene-based materials, and since it is known that graphene interacts weakly with most metals,[22] dispersion interactions were included through the DFT-D2 correction.[23] The plane waves' kinetic energy cut-off was 36 Ry, and the density cut-off was 576 Ry. Spin polarization was included in all the calculations. The first irreducible Brillouin zone was obtained using a *Γ*-centered 4×4×1 grid utilizing the general Monkhorst-Pack scheme,[24] while for the investigation of the electronic structure, a denser grid (20×20×1) was used.

Pristine graphene was modelled as $(3\sqrt{3}\times3\sqrt{3})R30^\circ$ supercell containing 54 carbon atoms, marked $C_{54}$. Doping graphene with boron was achieved by replacing $n$ carbon atoms with boron, resulting in models marked as $C_{54-n}B_n$, where $n$ is an integer from 1 to 3 (for $n$ = 0 this is pristine graphene). Boron concentration in these models is approx. 1.85 (for $n$ = 1), 3.70 (for $n$ = 2) and 5.56 (for $n$ = 3) atomic %. That way, we can also discuss the impact of boron concentration on the results obtained. For $n$ = 2 or 3, the optimal relative positions of B atoms were found and used for further investigations. Surface straining was modeled by applying biaxial strain along both the $x$ and $y$ axes. Four strain levels were imposed on the previously investigated B-doped graphene models: one compressive strain (−1%) and three tensile strains (+1%, +3% and +5%). The given levels of strains were chosen as the ones for which the planarity of the surface is conserved. Oxidized surfaces of pristine and boron-doped graphene were modelled by adsorption of an epoxy (O) or hydroxyl (OH) group on their preferential sites on the $C_{54-n}B_n$ surface. The models of $C_{54-n}B_n$ and their oxidized forms ($O@C_{54-n}B_n$ and $OH@C_{54-n}B_n$) were studied in detail in our previous work, ref. [25], and therefore will not be discussed here.

The strength of adsorption of the chosen metals (marked as M, which is Mg, Zn, Cu or Pt) on a given substrate is quantified by metal adsorption energy, $E_{\mathrm{ads}}(\mathrm{M})$, calculated as:

$$E_{\mathrm{ads}}(\mathrm{M}) = E(\mathrm{M@subs}) - [E(\mathrm{subs}) + E(\mathrm{M})] \tag{1}$$

where $E(\mathrm{M@subs})$, $E(\mathrm{subs})$ and $E(\mathrm{M})$ stand for ground state total energies of the system where M is adsorbed onto the substrate, of the bare substrate, and of the isolated metal atom, respectively. In this study, the substrate is pristine or boron-doped graphene, $C_{54-n}B_n$, their oxidized forms $O@C_{54-n}B_n$ or $OH@C_{54-n}B_n$, or their forms strained by $s$%: $C_{54-n}B_n(s\%)$, where $s \in \{-1, +1, +3, +5\}$. We note here that the metal adsorption energy can also be calculated with respect to metal atom in bulk metal phase, instead of isolated metal atom. The difference between the adsorption energy defined by eq. (1) and the one calculated with respect to metal atom in bulk metal would be a constant - the cohesive energy ($E_{\mathrm{coh}}$) of the given metal. Here, we report the energies calculated using eq. (1), but to reconcile the two possible definitions, and to assess metal aggregation, we compare them to cohesive energies of the studied metal later on in the discussion. While on pristine graphene there are only three inequivalent, highly coordinated possible adsorption sites (top, bridge, hollow), doping with B induces more sites that need to be probed for M adsorption, since the distance of the site from the dopant atom(s) also becomes a factor. The same happens upon adding O-groups to the systems – the surface is more "complicated" and therefore, more sites need to be probed. Here, we report only the most stable sites for each metal on each surface but note that a large total number of structures (starting adsorption sites) were tested. Adsorption energy was used as the primary stability descriptor, as it provides a direct measure of the thermodynamic preference for single-atom anchoring and is commonly employed as a screening criterion for single-atom catalysts. Comparison with the cohesive energy of the bulk metal further indicates the likelihood of atom dispersion versus aggregation, while kinetic effects such as diffusion barriers are beyond the scope of the present study.

Most investigated metals transfer some of their valence electrons to the substrate upon interaction, and therefore for the selected systems we have analyzed 3D charge difference ($\Delta\rho$) plots, where $\Delta\rho$ is calculated as:

$$\Delta\rho = \rho(\text{M@subs}) - [\rho(\text{M@subs}-\text{M}) + \rho(\text{M})] \quad (2)$$

where $\rho$(M@subs), $\rho$(M@subs−M) and $\rho$(M) stand for the ground state charge density of the system with the adsorbed metal, of the system M@subs from which M is removed (frozen geometry) and of isolated M.

In order to assess the suitability of single Pt atoms on $C_{54-n}B_n$ as potential electrocatalysts for hydrogen evolution reaction (HER), we have also investigated the adsorption of H onto Pt@$C_{54-n}B_n$, Pt@O@$C_{54-n}B_n$ and Pt@OH@$C_{54-n}B_n$ systems (substrates). Hydrogen adsorption energy was calculated with respect to ½ $H_2$ as:

$$E_{\text{ads}}(\tfrac{1}{2}\text{H}_2) = E(\text{H@Pt@subs}) - [E(\text{Pt@subs}) + \tfrac{1}{2}E(\text{H}_{2,\text{isol}})] \quad (3)$$

where $E$(H@Pt@subs), $E$(Pt@subs) and $E(H_{2,isol})$ stand for ground state total energies of the system where H is adsorbed onto the Pt atom at the chosen substrate, of the bare Pt@substrate system, and of the isolated hydrogen molecule, respectively. Gibbs free energy for H adsorption ($\Delta G$) was then assessed as:

$$\Delta G(\text{H}_{\text{ads}}) = E_{\text{ads}}(\tfrac{1}{2}\text{H}_2) + \Delta E_{\text{ZPE}} - T\Delta S(\text{H}) \quad (4)$$

where $\Delta E_{ZPE}$ is the difference in zero-point energy between the adsorbed hydrogen and hydrogen in the gas phase, $T$ is temperature and $\Delta S$(H) is the entropy difference between the adsorbed state and the gas phase. Following ref. [26], $\Delta G(H_{ads})$ is obtained by adding +0.24 eV to $E_{ads}(\frac{1}{2}H_2)$. For CO adsorption on the single Cu atom on (oxidized) boron-doped graphene, we use an equation analogous to the eq. (1) to calculate the adsorption energy of CO:

$$E_{\text{ads}}(\text{CO}) = E(\text{Co@Cu@subs}) - [E(\text{Cu@subs}) + E(\text{CO})] \quad (1)$$

where $E$(Co@Cu@subs), $E$(Cu@subs) and $E$(CO) stand for ground state total energies of the system where CO is adsorbed onto substrate containing a single Cu atom, of the Cu-substrate, and of the isolated CO molecule, respectively. Graphical presentations of all the structures were made using Vesta.[27]

## 3. Results and discussion

### 3.1. Metal adsorption onto non-oxidized boron-doped graphene

Pristine graphene ($C_{54}$) is found to interact with Mg and Zn, which preferentially occupy the $C_6$-hollow site (**Fig. 1**), rather weakly with corresponding metal (M) adsorption energies equal to −0.11 and −0.18 eV, respectively. The reason for such weak interactions can be found in the fully filled orbitals of Mg ([Ne] $3s^2$) and Zn ([Ar] $3d^{10}$ $4s^2$) and their relatively high ionization energies, in combination with the presence of the inert π-electronic cloud on both sides of pristine graphene. All this results in weak dispersion interactions and physisorption of Mg and Zn on pristine graphene. These metal atoms are hovering ≈ 3 Å above the corresponding hollow site (**Fig. 1**). Copper's interaction with pristine graphene is somewhat stronger, with the adsorption energy of −0.59 eV, occupying the C-top site (**Fig. 1**). On the other hand, Pt optimally binds to the C−C bridge site (**Fig. 1**), much more strongly than the first three metals, with $E_{ads}$ amounting to −2.05 eV. As expected, the interaction of pristine graphene is stronger with metals whose valence orbitals are not fully filled, Cu ([Ar] $3d^{10}$ $4s^1$) and especially Pt ([Xe] $4f^{14}$ $5d^9$ $6s^1$), because Pt's partially filled d and s orbitals hybridize with graphene's π system, leading to chemisorption. All the calculated metal adsorption energies and geometries on pristine graphene agree with previous literature reports,[22,28–31] with slight variations due to supercell sizes and chosen computational schemes.

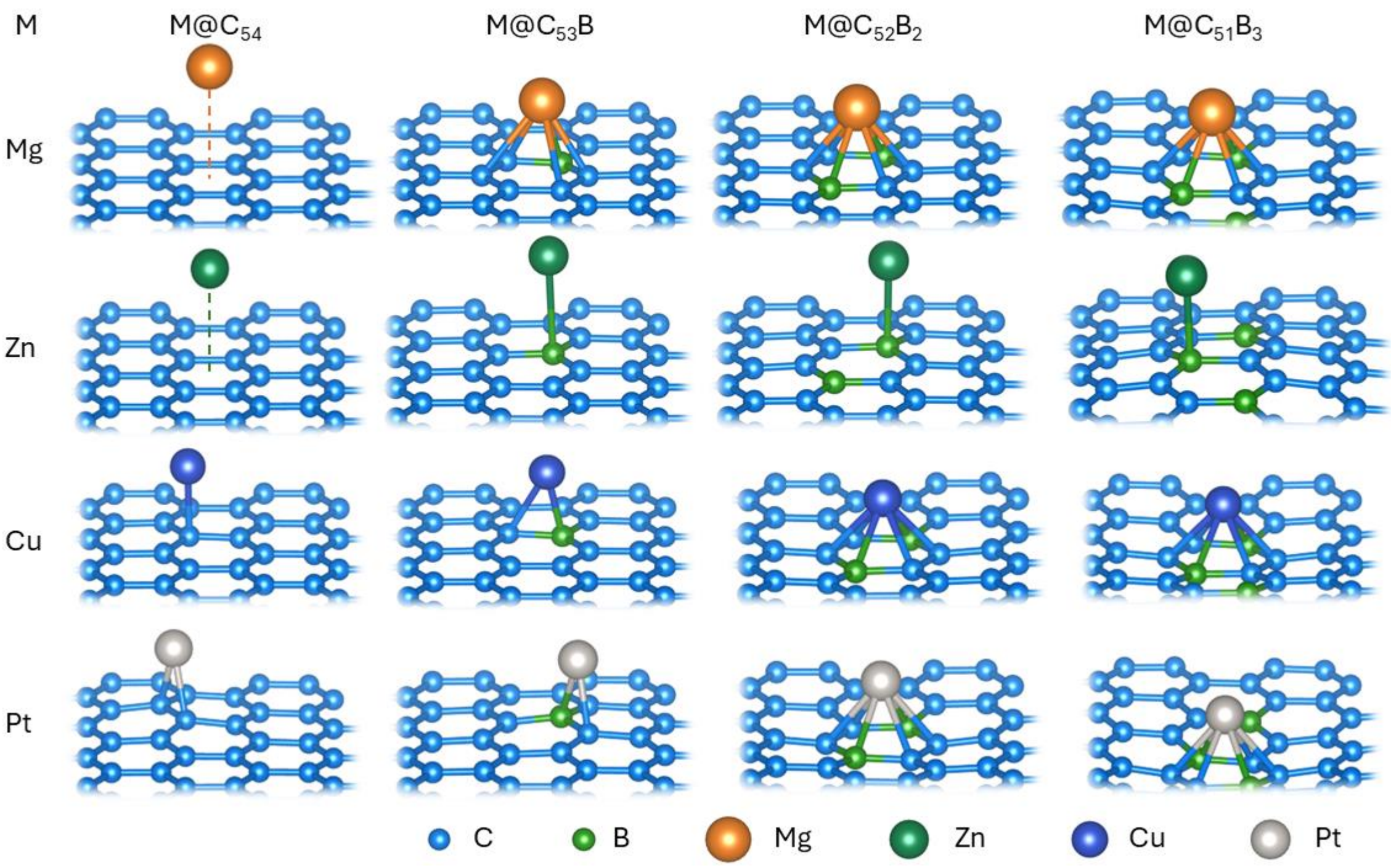

**Figure 1.** Optimized structures of chosen metal (M = Mg, Zn, Cu or Pt) adatoms on pristine ($C_{54}$) and B-doped graphene with various concentrations of B ($C_{53}B$, $C_{52}B_2$, $C_{51}B_3$), with a legend at the bottom indicating the elements represented by spheres of different sizes and colors.

Upon doping graphene with boron, we find that the preferential sites for metal adsorption change, as well as the corresponding adsorption energies. In general, we found that B-doping enhances the adsorption of the chosen metals. Substitutional doping by B, which has one valence electron fewer than C, makes the nearby sites electron-deficient, and creates a hole state (empty $p_z$). These sites are prone to accepting charge, which encourages metals to donate charge and bind more strongly near the B atom. In general, the B-induced empty $p_z$ states hybridize with the metal's valence orbitals more than those of the pristine C sites (Supplementary Information, Fig. SI1), so orbitals overlap improves and the bond localizes at B, B-adjacent bridge, or hollow sites. Thus, bonding becomes partially ionic instead of having a pure van der Waals physisorption character. For Mg and Cu, the strongest metal binding is achieved on graphene with approx. 3.70 at.% of boron, i.e. on the $C_{52}B_2$ model, with the adsorption energies amounting to −1.45 and −2.17 eV, respectively. Based on these two cases, it is obvious that the higher overall concentration of B does not guarantee stronger M binding. Instead, the local B motif (atomic arrangement) matters more, i.e. the proximity of more than one B to the adsorption site stabilizes M binding at that site, as the presence of more B atoms deepens the site's electron accepting character. Since, in the case of $C_{51}B_{3,}$ the third B atom is not found in the same hexagon as the first two, its influence on the adsorption energy of Mg and Cu is much smaller. The pronounced strengthening of the Cu binding implies that B-doping of graphene could also have great potential for removing Cu from industrial wastewater, as Cu is one of the metals commonly present in them and harmful to the environment [19]. For Zn and Pt, adsorption becomes stronger with each added boron atom, and so the strongest binding is found with approx. 5.56 at. % of boron, i.e. on the $C_{51}B_3$ model, with the adsorption energies of −0.43 and −3.40 eV, respectively. However, the relative change of $E_{ads}$ for these metals is much smaller when going from $C_{52}B_2$ to $C_{51}B_3$ than going from $C_{53}B$ to $C_{52}B_2$. Upon adding more boron dopant, Mg, Cu and Pt switch their preference to B-containing hollow sites, while Zn prefers binding to B-top (**Fig. 1**). The calculated adsorption energies are systematically shown and compared to metal cohesive energies (in order to assess the tendency for metal aggregation, which will be further discussed later on) in Supplementary Information, Table SI1. The observed shifts in preferred adsorption sites upon B-doping can be understood as a combined consequence of both electronic and geometric effects. From the electronic perspective, the presence of substitutional B atoms breaks the uniform π-electron distribution of pristine graphene and introduces localized electron-deficient regions. These regions modify the local potential energy landscape and favor adsorption geometries in which the metal atom can interact directly with the B-centered empty $p_z$ states. Consequently, adsorption configurations that maximize orbital overlap with these dopant-induced states (typically top or bridge sites adjacent to B, or B-containing hollow sites) become more stable than the high-symmetry hollow sites preferred on pristine graphene. In addition to electronic factors, geometric distortions induced by B incorporation also contribute to the change in adsorption preferences. The slightly larger covalent radius and different bonding characteristics of boron relative to carbon introduce local lattice strain. These structural perturbations reduce the equivalence of adsorption sites and can sterically favor specific metal positions depending on the particular B arrangement. For multi-boron models, the exact spatial distribution of dopants determines whether top, bridge, or hollow configurations provide the most favorable coordination environment. The relative importance of these effects varies with the nature of the metal. Cu and Mg, which more readily donate electron density, are particularly sensitive to the electron-accepting character of B sites, leading to pronounced shifts toward B-adjacent hollow positions. In contrast, Zn, with a more filled d-shell and weaker tendency for charge transfer, preferentially binds directly on top of the B atom where electrostatic stabilization dominates. Pt, owing to its more delocalized d-states and stronger covalent interaction with the carbon framework, exhibits a smaller change in site preference, although stabilization near B-containing hollows is still observed.Bader analysis of the studied metal@substrate systems reveals that nearly linear correlation between the metal adsorption energy and the charge transferred from the metal atom onto the substrate ($\Delta q$(M)) exists for Mg, Zn and Cu binding onto $C_{54-n}B_n$, while for the case of Pt the corresponding coefficient of determination for linear correlation is much lower (**Fig. 2**). The observed linear correlation for Mg and Zn indicates that electron donation is the main driving factor for the adsorption strength. The case of Cu, with $R^2$=0.92, can also be considered as mostly charge-transfer driven interaction, but with some additional effects. When the correlation is not linear, as for Pt, it implies that orbital hybridization dominates. Pt has accessible d and s states that hybridize with the graphene's π states that leads to chemisorption even without the presence of the dopants. Non-linear correlation between $E_{ads}$(M) and $\Delta q$(M) could also imply that on different substrates different surface atoms' orbitals are responsible for the metal binding. This could be a result of the change of the preferential adsorption site (we do see that for Cu and Pt in Fig. 1: for example, Pt prefers the two-coordinated bridge on $C_{54}$ and $C_{53}B$, but the six-membered hollow sites on $C_{52}B_2$ and $C_{51}B_3$).

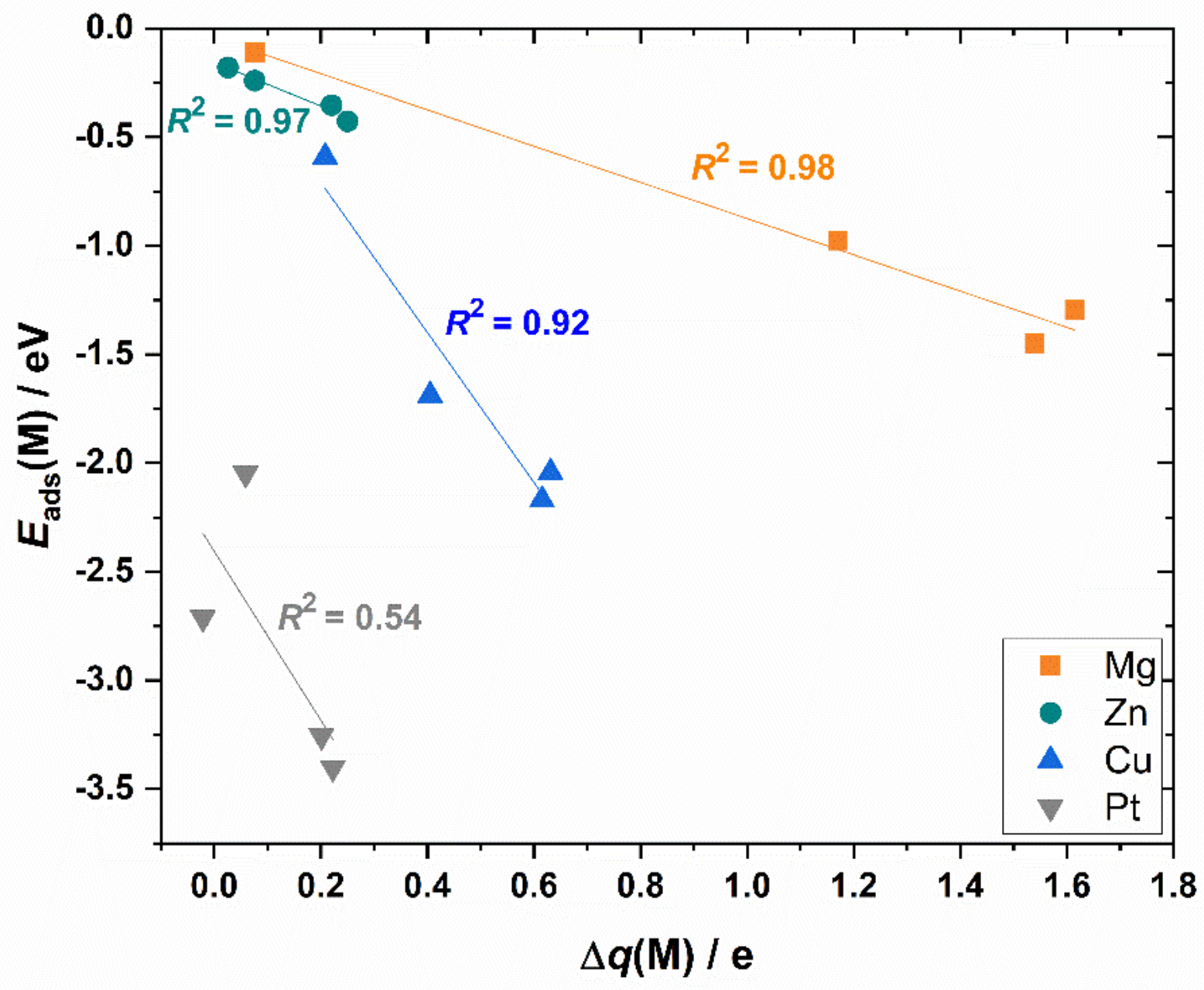

**Figure 2.** Correlation between the metal adsorption energies ($E_{ads}$(M)) and charge transferred from the metal atom onto the substrate (Δ$q$(M)), upon metal adsorption onto $C_{54-n}B_n$, where $n$ is an integer from 0 to 3. Corresponding coefficients of determination ($R^2$) are shown for each metal.

Next, we address the impact of mechanical strain, which is known to be a powerful parameter in tuning the electronic structure and catalytic performance of two-dimensional materials. We have applied four levels of biaxial strain on the previously investigated B-graphene models: compressive strain (−1%) and three levels of tensile strain (+1, +3 and +5%). Upon such levels of strain, the B-graphene layer is found to remain planar, so we aimed to exclude the effects of surface corrugation. The changes in the electronic structure of the $C_{54-n}B_n$ models caused by the strain are shown in Supplementary information, Fig. SI2. The obtained adsorption energies (Fig. 3) reveal that the chosen levels of strain have relatively small impact on the strength of metal interaction with B-graphene.

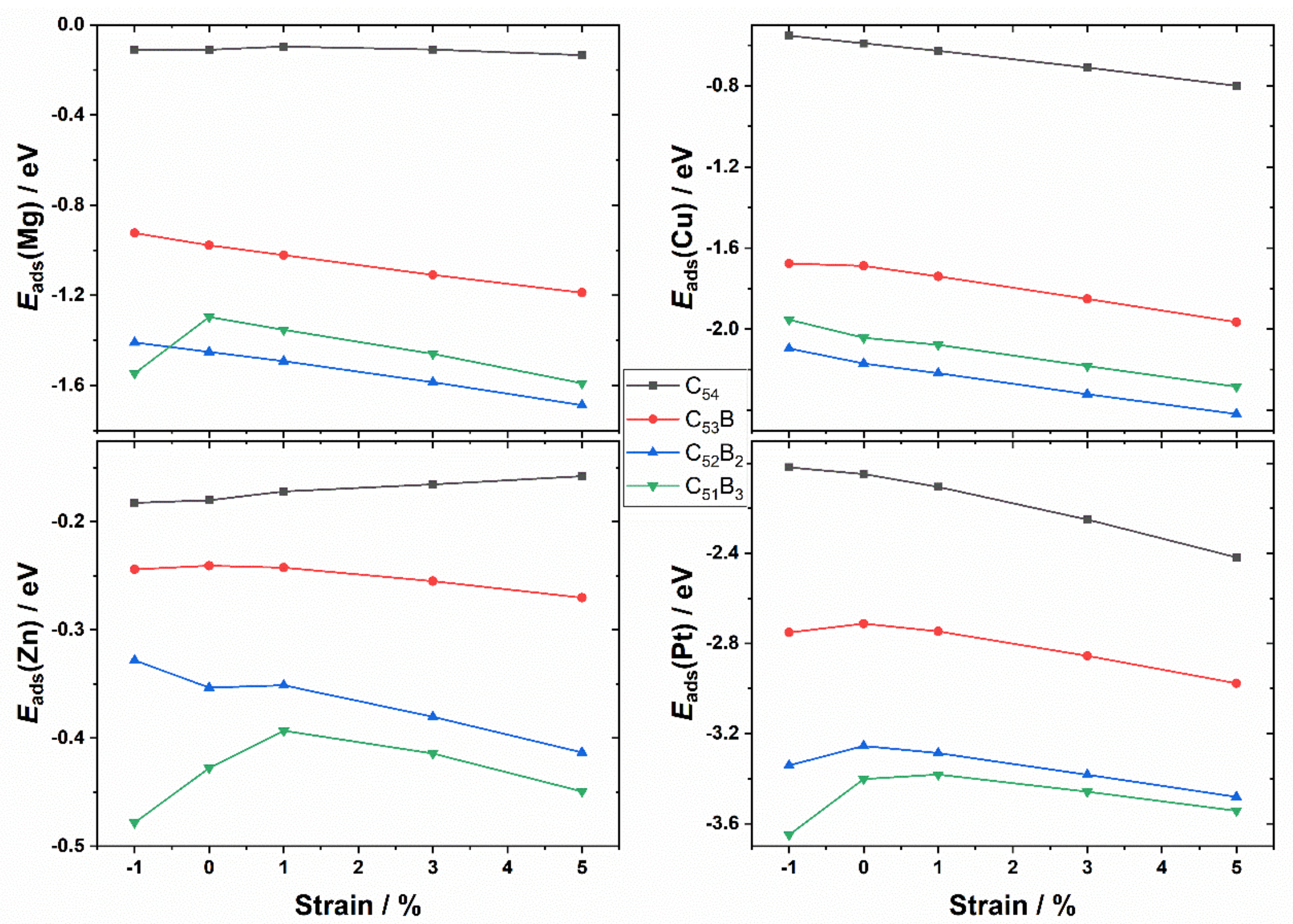

**Figure 3.** Effects of strain on the adsorption energies of Mg, Cu, Zn and Pt onto $C_{54-n}B_n$ models (0% on the *x*-axis represents non-strained $C_{54-n}B_n$ surfaces).

Overall, strain (from −1 to + 5%) alters the adsorption energies of the chosen metals by max. 0.37 eV on pristine graphene and by max. 0.29 eV on B-doped graphene, i.e. the effects of strain are smaller when graphene is B-doped. Based on these values, it could be said that strain only fine-tunes the adsorption strength of the metals on B-doped graphene. That is somewhat expected, since the electronic structure of the surface does not change significantly in the region from −2 to 0 eV (*vs*. Fermi energy) upon applying strain (Supplementary Information, Fig. SI2). The states in that energy region are generally thought to be mostly responsible for the reactivity of the graphene-based materials.[32] However, a slight alteration of the adsorption energy is not the only effect of applying strain on B-doped graphene; in some cases, we find that the compressed surfaces corrugate upon the interaction with the metal. For example, looking at Fig. 3, we see clear downward trends for Mg adsorption on strained B-doped graphene – tensile strain makes the interaction somewhat stronger, while compression makes it a bit weaker, but with one prominent exception: Mg on the −1% (compressed) $C_{51}B_3$ surface. In that case, we find that Mg adsorption induces corrugation of the B-doped surface (Fig. 4), and therefore the calculated adsorption energy does not reflect only adsorption, but also the deformation of the substrate. A similar effect is found for Zn and Pt on the same surface, which remains planar only upon Cu adsorption. Therefore, in the case of Cu, we do not observe a deviation from the trendline at $C_{51}B_3$(−1%) as we do for Mg, Zn and Pt. Under compressive strain, the Cu@$C_{51}B_3$ system accommodates the strain primarily through in-plane lattice distortion rather than out-of-plane buckling, since the Cu/B-graphene interaction does not provide sufficient driving force to locally deform the carbon network. On the other hand, for Mg, Zn and Pt local out-of-plane distortion (surface corrugation) becomes energetically favorable under compressive strain, as it relieves in-plane stress while preserving optimal M−C bonding distances. For Pt, this effect is further enhanced by significant d-state hybridization, leading to the strongest local structural response. It is noteworthy that such a corrugation also happens upon Pt adsorption onto compressed $C_{52}B_2$ and $C_{53}B$ (Supplementary Information, Fig. SI3), but not with other investigated metals. In case of strong interactions, like with Pt, higher B concentration is found to make the material more prone to corrugation upon metal adsorption under small compressive strain. In order to assess the contribution of the deformation energy of the substrate to the metal adsorption energies, we have calculated the energies of the "frozen" $C_{51}B_3$(−1%) systems obtained from the optimal geometries of M@$C_{51}B_3$(−1%), M = Mg, Zn, or Pt, by removing the metal atom and fixing the substrate in the corrugated shape (shown in Fig. 4). We found that the deformation energies ($E_{def}$, calculated as the difference between the corrugated and the flat surface) are up to 0.23 eV. If Mg, Zn and Pt adsorption energies in Fig. 3 were shifted up by the values of $E_{def}$, the observed exceptions from the trends would be much less prominent.

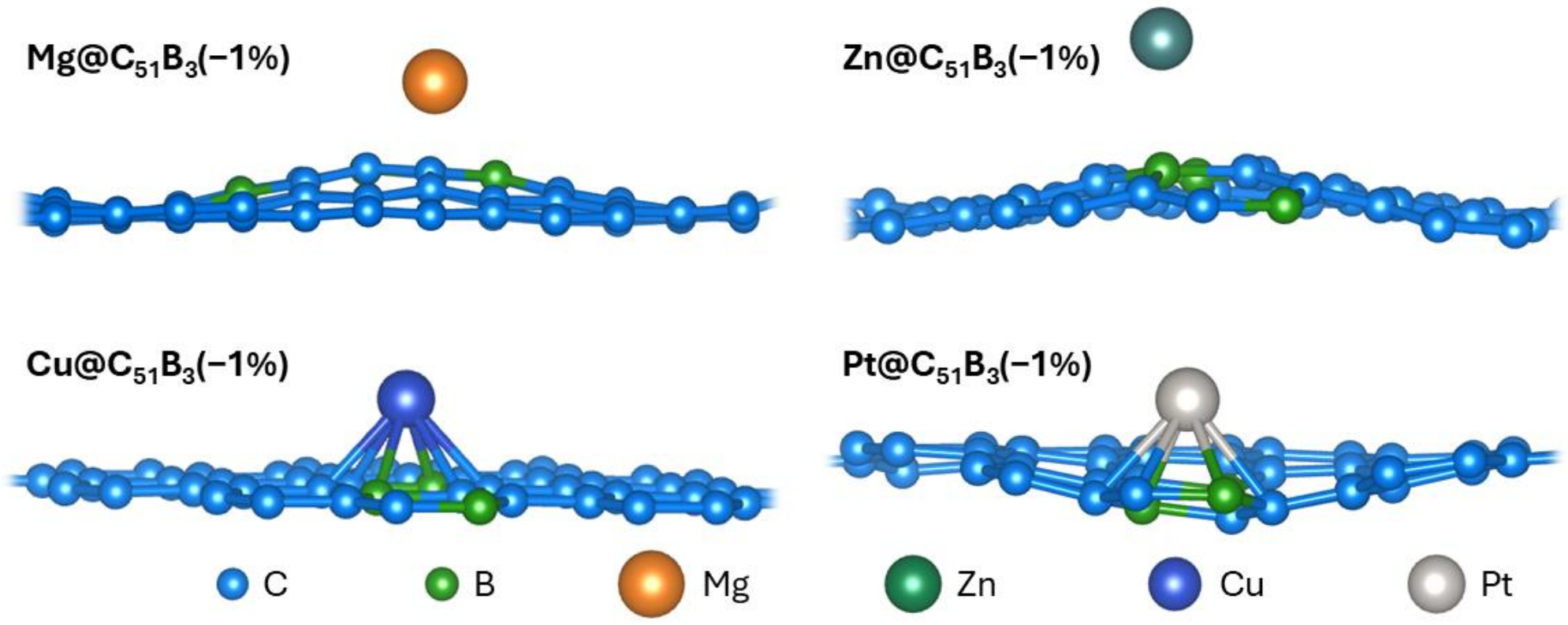

Figure 4. Optimized structures of M@$C_{51}B_3$(−1%), with a legend at the bottom indicating the elements represented by spheres of different sizes and colors.

Although the strain-induced variations in adsorption energies are relatively modest compared to the effects of boron doping, differences on the order of 0.1-0.3 eV can still be chemically meaningful in practical contexts. At ambient temperatures, an energy change of ~0.1 eV corresponds to a several-fold change in equilibrium constants and surface coverages. Therefore, even small shifts in adsorption strength may influence the stability of isolated metal atoms, their resistance to migration or aggregation, and the relative population of competing adsorption sites. In single-atom catalysis, such changes can affect reaction barriers and turnover frequencies, particularly for processes that are sensitive to the precise metal-support interaction strength. The applied strain range (−1% to +5%) is within experimentally accessible limits for graphene and graphene-based systems. Graphene is known for its exceptional mechanical strength and flexibility, with experimentally demonstrated elastic strain well above 5% before mechanical failure.[33] In practice, strain levels of a few percent are routinely achieved using several approaches, including flexible or stretchable substrates, thermal expansion mismatch, substrate engineering, and mechanical bending or stretching.

3.2. Metal adsorption onto oxygen-functionalized boron-doped graphene

Oxidation of graphene is known to alter its reactivity and the ability to bind metal atoms.[19,25,34] Therefore, we also investigate the adsorption of the chosen metals onto epoxy- and hydroxyl-functionalized non-doped and B-doped graphene. Oxygen functional groups on graphene tend to aggregate into oxidized patches rather than being uniformly distributed.[35] Such a morphology results in locally functionalized regions separated by relatively pristine areas, particularly at low to moderate oxidation levels. To model the fundamental interaction mechanism, a single functional group was placed in a sufficiently large supercell, corresponding to the low-coverage limit and minimizing interactions between periodic images. This approach allows the adsorption behavior to be analyzed in terms of the local chemical environment, while avoiding additional cooperative effects from neighboring functional groups. Although increased functional group density within oxidation domains may modify absolute adsorption energies, the site preference and relative adsorption trends are expected to remain qualitatively robust.

The epoxy group on non-doped graphene is found to interact directly with Mg and Cu, resulting in MgO or CuO species adsorbed onto graphene – the epoxy group is “swept” away from the basal plane by these metals (**Fig. 5**). However, once B is present in graphene, the epoxy group is attached to the surface more strongly [25] and interacts with Mg and Cu without any surface reduction, i.e. the presence of boron stabilizes the interaction. The trend in the adsorption energies of Mg regarding B concentration is the same as for the non-oxidized systems. The adsorption energies of Mg are up to ~0.5 eV more negative (stronger binding) than on the corresponding surfaces without epoxy groups. The strongest Mg, Cu and Zn adsorption is observed on epoxy-graphene with approx. 3.70 at.% of B (M@$C_{52}B_2O$), with the corresponding $E_{ads}$ amounting to −1.87, −2.06 and −0.50 eV, respectively. However, for Cu and Zn the adsorption energies on $C_{54-n}B_nO$ are not always more negative than those on corresponding $C_{54-n}B_n$ but are within 0.15 eV from them. It is interesting to note that when B is present, Pt prefers interacting with epoxy boron-doped graphene by clinging to the opposite side of the plane than O. When B is introduced into the graphene lattice, the preferred geometry for Pt on epoxy-graphene changes and Pt stabilizes on the side opposite to the epoxy group. This behavior can be attributed to the electron-deficient character of substitutional B, which induces local charge redistribution and activates the neighboring B–C framework toward metal binding. In this case, the Pt–support interaction with the B-containing region becomes energetically more favorable than the Pt–O interaction. Adsorption on the opposite side enables stronger Pt–C/B bonding while avoiding steric hindrance and electronic competition with the oxygen atom. Obviously, this behavior is metal-dependent. While Pt prefers adsorption on the side opposite to the epoxy group in the B-doped system, Mg, Cu, and Zn remain stabilized on the same side as the oxygen atom. This difference originates from the nature of the metal/substrate interaction. Mg and Zn interact with graphene predominantly through electrostatic and weak donor/acceptor interactions, while Cu exhibits intermediate binding strength. For these metals, the additional stabilization provided by the electronegative oxygen atom (through local charge polarization and metal/O interaction) outweighs the benefit of stronger coupling to the B-doped lattice. The strongest Pt adsorption is found on $C_{51}B_3O$, with $E_{ads}$ equal to −3.53 eV. The calculated adsorption energies are systematically shown and compared to metal cohesive energies (to assess the tendency for metal aggregation, which will be discussed later on) in Supplementary Information, Table SI2.

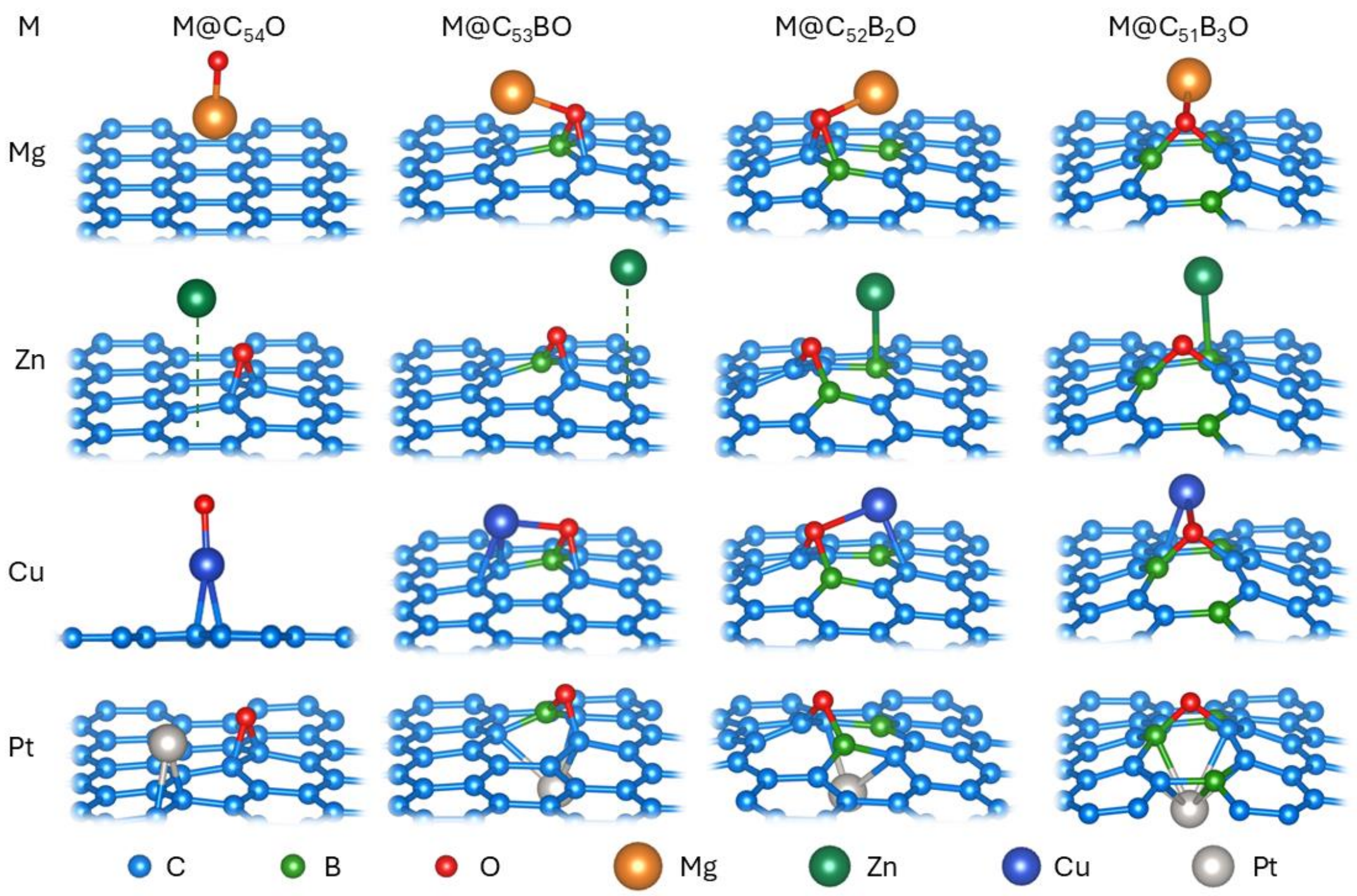

**Figure 5.** Optimized structures of chosen metal adatoms on epoxy-functionalized pristine and B-doped graphene, $C_{54-n}B_nO$, where $n$ is an integer from 0 to 3, with a legend at the bottom indicating the elements represented by spheres of different sizes and colors.

When pristine or B-graphene is functionalized (oxidized) with hydroxyl instead of epoxy group, the nature and strength of the interactions with a metal change. To begin with, regardless of the B concentration, Mg interacts directly and very strongly with the OH group, resulting in phase separation, i.e. OH detachment from the surface (Fig. 6, top row). Such strong Mg−O interaction suggests a trade-off between anchoring strength and structural stability, which in practical systems may be balanced by controlling oxygen coverage or combining oxygen functionalities with heteroatom dopants. Similar happens with Cu, Zn and Pt upon interaction with OH on non-doped graphene, where OH is "swept" from the basal plane. Regarding the interactions of Cu, Zn and Pt on hydroxyl-functionalized B-doped graphene, they are found to be stable in the sense that OH remains on the surface. This is a direct consequence of the fact that a single OH group is approx. 3 times more strongly bound to B-doped compared to pristine graphene.[25] For Cu, Zn and Pt on hydroxyl-functionalized B-doped graphene, we observe prominent interaction between the metals and O from OH, but also some metal interaction with the basal plane (Fig. 6). The adsorption of Cu, Zn and Pt is getting stronger with increasing concentration of boron in graphene, with the most negative metal adsorption energies on B-doped graphene with approx. 5.56 at.% of B (model $C_{51}B_3OH$) amounting to −3.06, −1.11 and −3.42 eV, respectively. The calculated metal adsorption energies onto OH-functionalized B-doped graphene are systematically shown and compared to metal cohesive energies (to assess the tendency for metal aggregation, which will be discussed later on) in Supplementary Information, Table SI3. It should be noted that under operating electrochemical conditions CuOH species may interact competitively with the aqueous environment. A quantitative assessment of CuOH stability would therefore require comparison of adsorption on B-doped graphene with solvation and dissolution energies in water, which is beyond the scope of the present work. Nevertheless, the substantially stronger binding of CuOH on B-doped graphene relative to pristine graphene indicates an increased resistance to desorption or leaching.

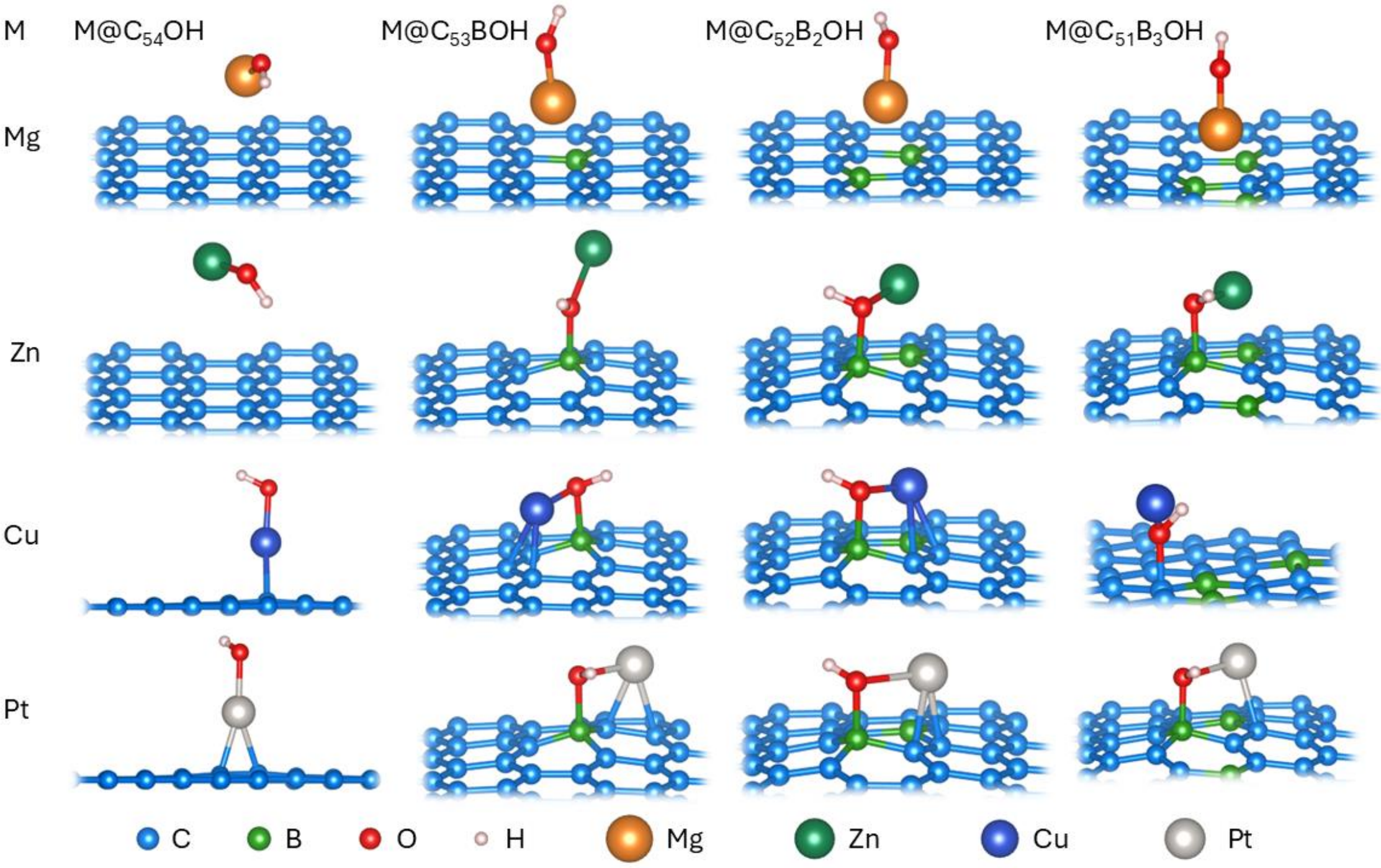

**Figure 6.** Optimized structures of chosen metal adatoms on hydroxyl-functionalized pristine and B-doped graphene, $C_{54-n}B_nOH$, where *n* is an integer from 0 to 3, with a legend at the bottom indicating the elements represented by spheres of different sizes and colors.

An overview of the effects of boron concentration and surface oxidation on the adsorption energies of the chosen metals is shown in Fig. 7. It should be noted that for the cases where a phase separation occurs, the calculated adsorption energies do not reflect the real strength of metal adsorption, since they also include new phase formation. Enhanced Cu and Zn binding on B-graphene could be interesting from the viewpoint of removing these metals from industrial wastewater in which they are often present.[19] From the viewpoint of metal-ion batteries, B-graphene shows a greater potential compared to pristine graphene due to enhanced interaction with these metals. However, Mg−OH separation could be problematic in the case of oxidized B-graphene, since it would induce irreversible changes of the electrode material.

3.3. Potential applications – metal anchoring

The simplest criterion for deciding if a chosen substrate could be suitable for anchoring chosen metal atoms (e.g. in metal-ion batteries or SACs) is comparing $E_{ads}(M)$ and the cohesive energy ($E_{coh}(M)$) of the given metals. Taking absolute values, if $|E_{ads}(M)|$ is larger than $E_{coh}(M)$, that would indicate that for M it is energetically more favorable to be on this substrate than in its pure metallic phase, i.e. that in the case of high M concentration, M atoms would prefer clinging to the substrate rather than cluster together. With that idea, we have compared the calculated $|E_{ads}(M)|$ with $E_{coh}(M)$. The cohesive energies of the chosen metals are 1.51 eV per Mg atom, 1.35 eV per Zn atom, 3.49 eV per Cu atom and 5.84 eV per Pt atom.[36] As Fig. 7 shows, $|E_{ads}(M)| > E_{coh}(M)$ is true only for Mg@$C_{52}B_2O$ among the investigated systems (we exclude the systems for which phase separation occurs). Therefore, we took a closer look at $Mg@C_{52}B_2$ and its oxidized forms.

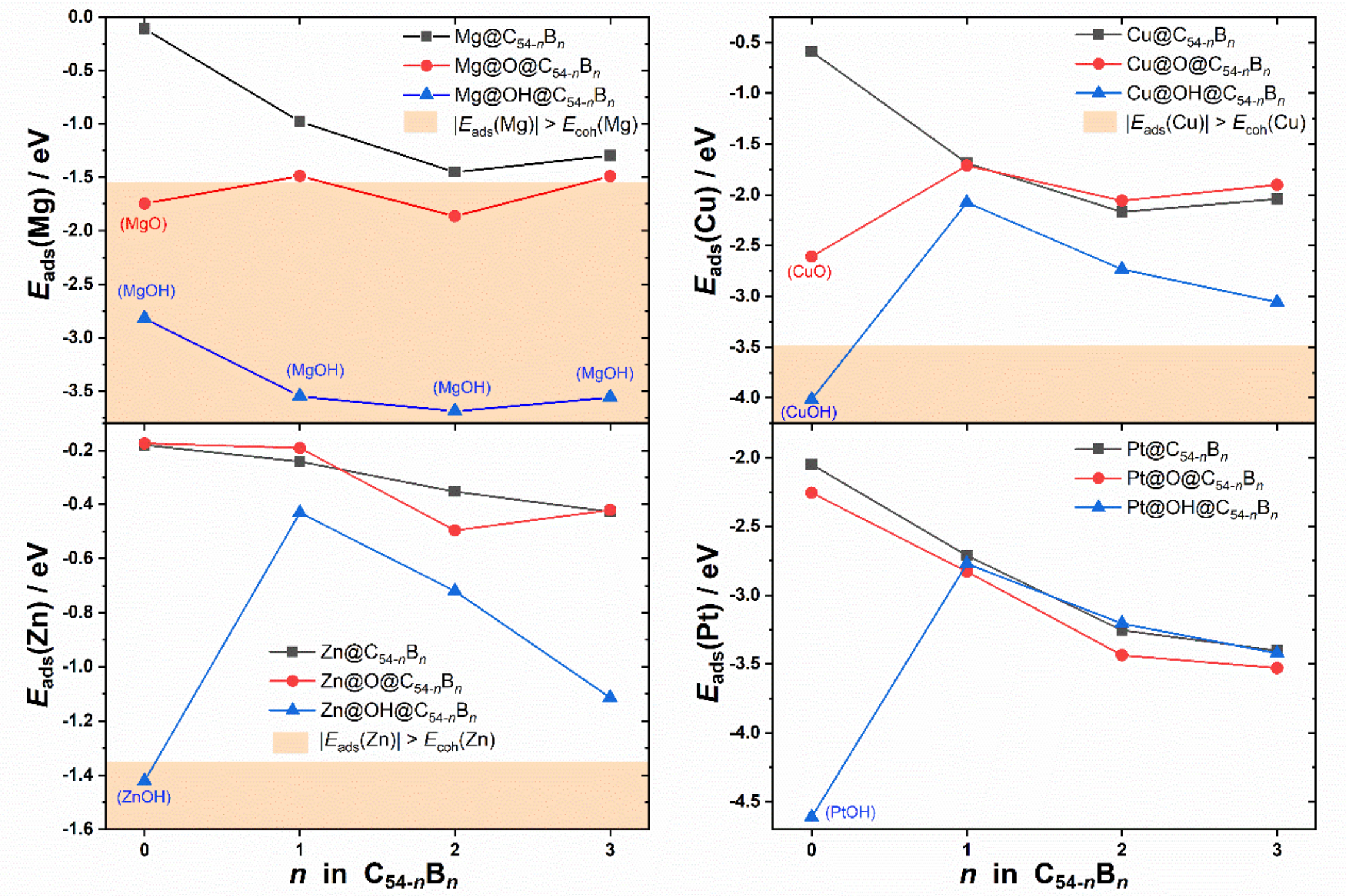

**Figure 7.** Adsorption energies of chosen metal atoms (Mg, Zn, Cu and Pt) on their preferential sites on $C_{54-n}B_n$ surfaces, as well as on the oxidized $O@C_{54-n}B_n$ and $OH@C_{54-n}B_n$ surfaces ($n \in \{0, 1, 2, 3\}$). Colored areas of the graphs show the regions of stability for each metal, where the cohesive energies would be overcome by $E_{ads}$ (Pt is excluded, since it is way below, at 5.84 eV). Species in the parenthesis mark phase separation in the given cases.

For Mg adsorption on $C_{52}B_2$ and $C_{52}B_2O$, we observe a clear difference in the nature of Mg binding compared with the optimized structures discussed above (**Figs. 1** and **5**). This can also be confirmed by the DOS analysis of the mentioned systems. In the case of $Mg@C_{52}B_2$, magnesium is situated at the $C_4B_2$-hollow site (**Fig. 1**), where it interacts equally with both B atoms, and therefore B(p) states of both B atoms are identical (**Fig. 8**). A very sharp Mg (s) peak appears just below and just above Fermi energy, which is typical of weakly interacting s-type adatom. Slight redistribution of C (p) and B (p) density of states occurs, but overall, the underlying substrate DOS remains similar to that of clean $C_{52}B_2$. When it comes to Mg adsorption onto $C_{52}B_2O$, the situation is different due to the presence of the epoxy group on B-doped graphene. In the bare $C_{52}B_2O$ surface, O is bound to C−B bridge, which results in a significant overlap of oxygen's p states with the p states of C and B forming that bridge site (**Fig. 8**, top right). When Mg is added to the system, it interacts mostly with O and, as expected, one observes the Mg states above the Fermi energy (**Fig. 8**, bottom right). The analysis of Bader charges reveals that approx. 1.5 e is transferred from Mg to $C_{52}B_2$, while all 2 e all transferred from Mg to $C_{52}B_2O$. Charge difference plots for Mg on $C_{52}B_2$ and its oxidized forms confirm that Mg behaves as an electron donor in all cases, with charge depletion regions located predominantly around Mg atom (**Fig. 9**). The degree and directionality of charge transfer are strongly dependent on the presence and type of the surface functional group. The O-functionalization creates a strong Mg–O bond with significant charge flow and substrate involvement.

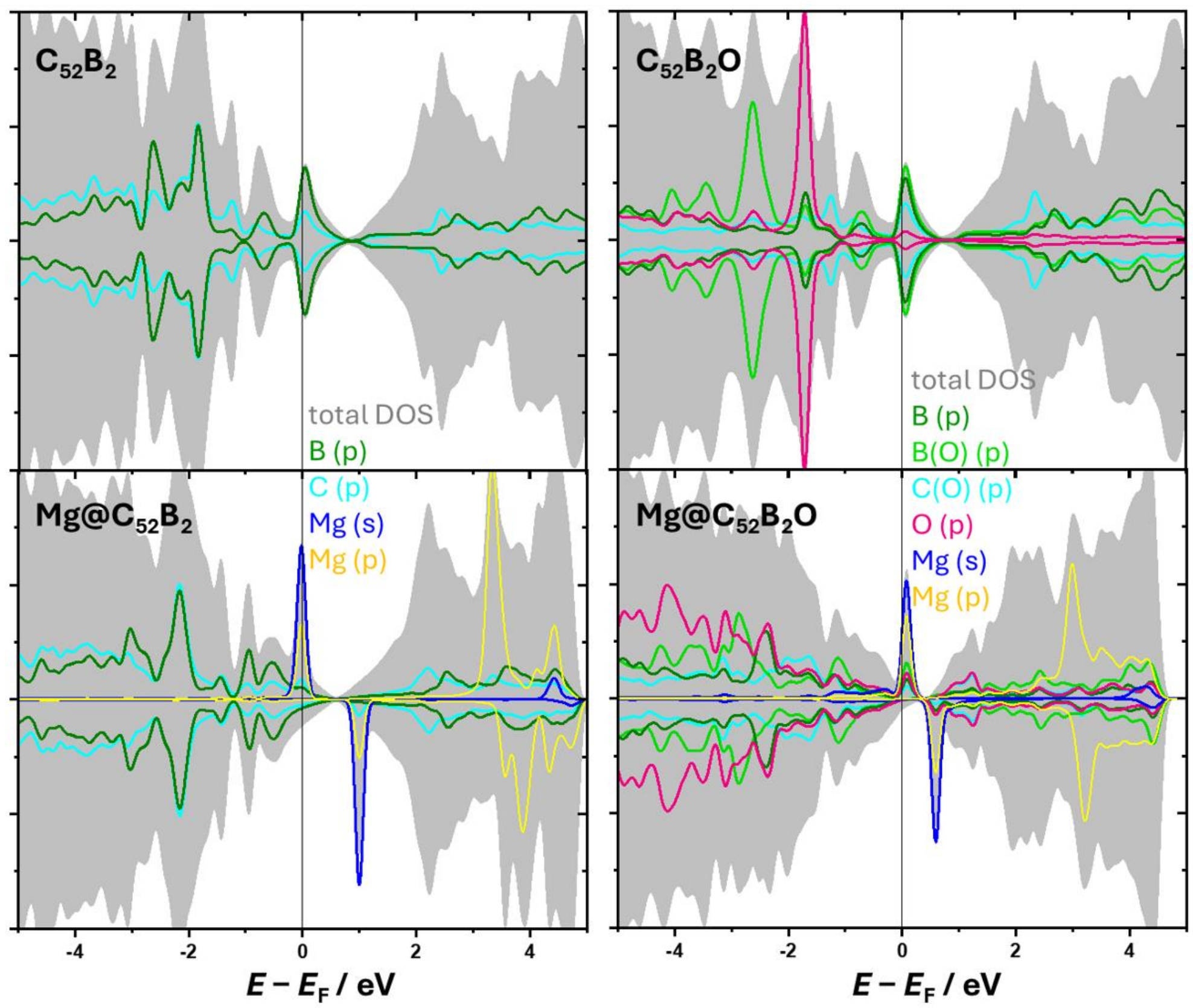

**Figure 8.** Densities of states (DOS) for non-oxidized $C_{52}B_2$ (left) and epoxy-functionalized $C_{52}B_2O$ (right), before (top) and upon (bottom) Mg adsorption. Total DOSes of the whole systems are given by gray shaded areas, while the colored lines represent orbitals of atoms of importance for adsorption. On the left, p orbitals of C and B atoms which form $C_4B_2$-hollow site for Mg adsorption are shown (states of both B atoms are the same). In case of oxidized $C_{52}B_2O$ (right), we show p orbitals of C and B atoms which form the bridge for O (marked as C(O) and B(O)), as well as of the second B atom. In the bottom DOSes, Mg's s orbitals are also shown. Total DOS is divided by a constant for a clearer view.

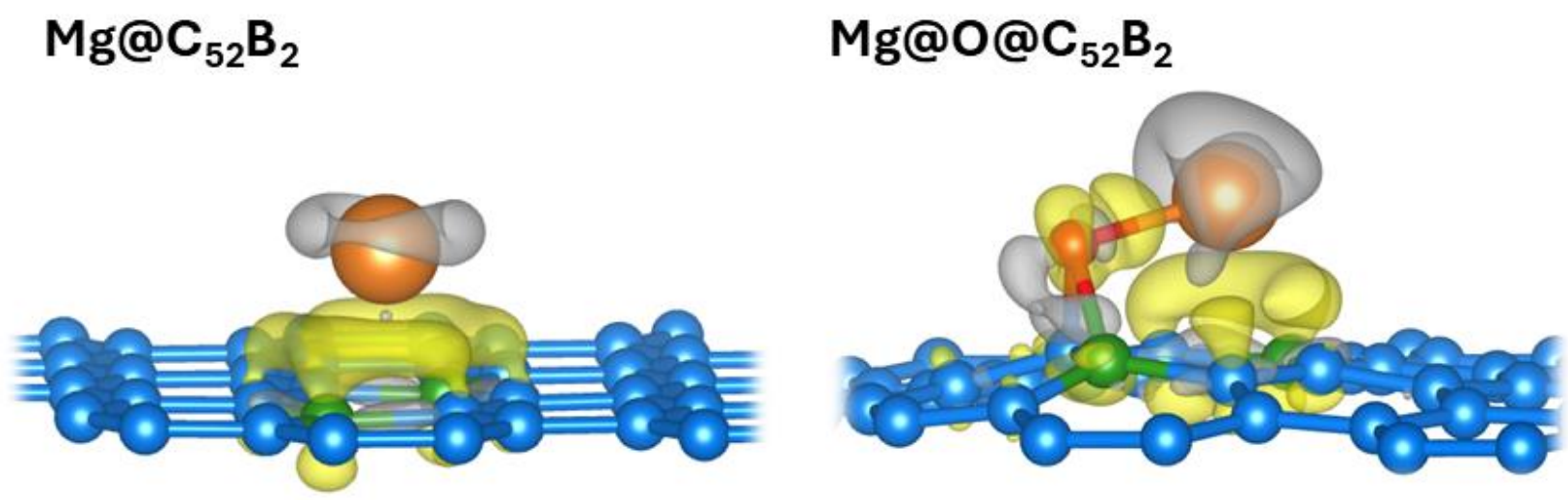

**Figure 9.** Charge difference plots (calculated using eq. 2) for Mg adsorption onto non-oxidized (left), and epoxy functionalized (right) B-doped graphene $C_{52}B_2$. Yellow isosurfaces indicate charge gain, while the grey ones indicate charge loss.

Even though the cohesive energy is overcome only for Mg adsorption on $C_{52}B_2$ and $C_{52}B_2O$, we do not fully dismiss the other systems, as we are aware of the "overbinding tradeoff" – very strong metal anchoring can impede surface diffusion of the metal, so it can make its uniform deposition on the surface (during the preparation of the catalyst) hard to achieve. In that spirit, even though the adsorption energies of Pt and Cu on the investigated substrates do not overcome their cohesive energies, we were still interested in probing the activity of the single Pt and Cu atoms as model SACs. We did this by testing the reactivity of single Pt atom on (oxidized) B-doped graphene towards atomic hydrogen - the intermediate in Hydrogen Evolution Reaction (HER), while for Cu we probed its reactivity towards CO, since Cu-based materials remain the best metal electrocatalysts for $CO_2$ reduction ($CO_2R$) towards generating hydrocarbon products (Introduction). CO binding strength was proposed as a descriptor for $CO_2R$ activity due to its linear correlation with reaction barriers of the $CO_{ads}$ reduction step on multiple metallic surfaces.[17] We probed H adsorption on top of Pt on $C_{54-n}B_n$ and their oxidized forms, $C_{54-n}B_nO$ and $C_{54-n}B_nOH$. The obtained adsorption energies of H ranged between −2.71 and −3.07 eV, i.e. adsorption energies of ½$H_2$ between −0.47 and −0.82 eV (calculated using eq. 3), resulting in $\Delta G(H_{ads})$ (eq. 4) ranging between −0.23 and −0.58 eV (Supplementary Information, Table SI4). Since the ideal HER electrocatalyst should have $\Delta G(H_{ads})$ close to zero,[26] and considering that the error of DFT calculated adsorption energies is about ±0.2 eV,[37] our best candidates are Pt on $C_{54}O$, $C_{53}BOH$ and $C_{51}B_3OH$. When it comes to CO adsorption on Cu at $C_{54-n}B_n$ and their oxidized forms, we find that CO binds relatively strongly to these surfaces, with adsorption energies between approx. −1.6 and −2.1 eV (Supplementary Information, Table SI5). We note that the presence of the epoxy or hydroxyl group on the surface does not affect CO adsorption significantly – CO binds directly to Cu in all the cases, by forming a Cu−C bond approx. 1.78 Å long, and the adsorption energies are all within 0.5 eV. Zhong *et al.* identified the optimal CO adsorption energy for $CO_2R$ to hydrocarbons on metal catalysts to be −0.67 eV.[38] Taking that value into account, Cu at $C_{54-n}B_n$ and their oxidized forms would be considered to bind CO too strongly, and therefore probably not be good candidate for $CO_2R$ electrocatalysts. Other than CO adsorption, for $CO_2RR$ applications the adsorption of $CO_2$ onto the surface is obviously important. The calculated adsorption energies (−0.30 to −0.62 eV, Supplementary Information, Table SI6, bond lengths and angles are also shown) fall within the moderate binding regime generally considered favorable for $CO_2$ reduction,[39] which may provide sufficient surface residence time for activation without risking site blocking by strongly bound species. We find that $CO_2$ molecule is activated (bent) on Cu at non-oxidized surfaces (Cu@$C_{54-n}B_n$), and on Cu@$C_{53}BO$ (**Fig. 10**). On other Cu at oxidized B-graphene surfaces, adsorbed CO2 remains nearly linear. Regarding adsorption energies, epoxy-functionalized B-doped graphene shows the strongest interaction and is therefore expected to enhance local $CO_2$ concentration, whereas OH functionalization has a weaker effect. According to the adsorption energies and activation (bending) of $CO_2$, Cu at low concentration epoxy-functionalized B-doped graphene seems to be the best candidate for $CO_2RR$. Obviously, B and O concentrations need to be carefully tailored to maximize the performance of potential catalytic material.

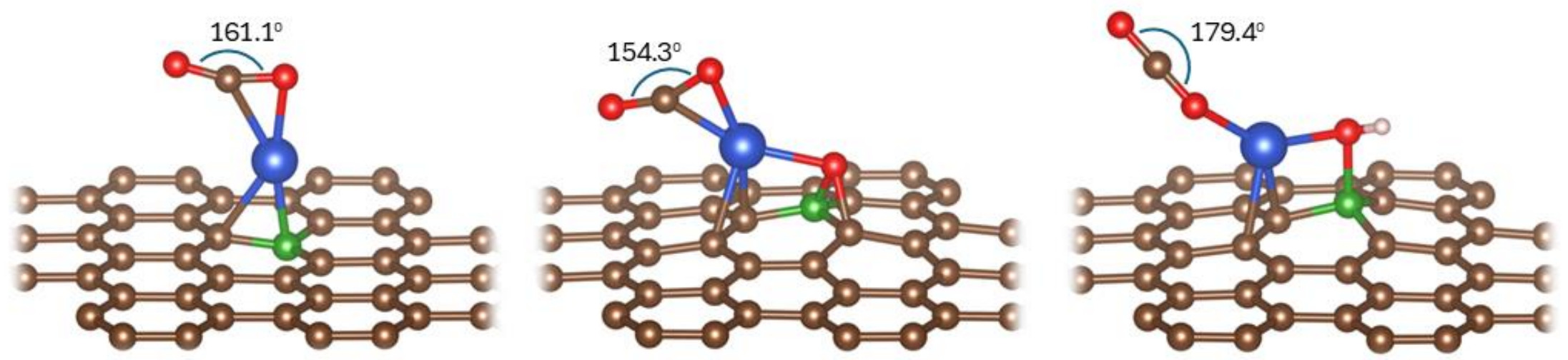

**Figure 10.** Optimized structures of $CO_2$ interacting with Cu@$C_{53}$B (left), Cu@$C_{53}$BO (middle) and Cu@$C_{53}$BOH (right), with O−C−O angles indicated for each.

Overall, the results indicate a trade-off between efficient reactant capture and potential inhibition by strongly bound intermediates, highlighting the importance of tuning the local chemical environment to balance adsorption strength and catalytic turnover. While the relatively strong binding of CO on Cu supported B-doped graphene (−1.6 to −2.1 eV) indicates that these systems are unlikely to be optimal for $CO_2$ electroreduction, where weaker CO adsorption is typically required to facilitate product desorption, such strong CO interaction may be advantageous for catalytic processes involving CO activation, such as CO oxidation. In this context, the calculated adsorption energies are significantly more negative than those reported for other single-atom catalysts proposed for CO oxidation (e.g., −0.917 eV for Y-doped $Ti_2CO_2$ (MXene) monolayer [40]). This suggests that Cu atoms stabilized on B-doped graphene could be promising candidates for heterogeneous CO oxidation catalysis.

Finally, we note that for applications involving metal capture or immobilization, the dominant factor is the overall magnitude of adsorption energy, which is here primarily governed by B-doping. Nevertheless, strain can provide an additional means of fine-tuning selectivity among different metals or optimizing binding under mechanically deformed or flexible graphene supports. Furthermore, the strain-induced corrugation observed for some compressed systems may alter the local coordination environment of the adsorbed metal, potentially affecting catalytic behavior beyond what is reflected by adsorption energies alone. Additionally, a quantitative assessment of migration barriers or clustering energies could provide further valuable complementary insight into the dynamic behavior of single atoms under operating conditions.

## 4. Conclusion

In this work we have systematically investigated how boron doping, mechanical strain, and surface oxidation influence the adsorption behavior of Mg, Zn, Cu and Pt on graphene, which are of interest for electrochemical energy conversion applications. Pristine graphene interacts only weakly with Mg and Zn and moderately with Cu, while Pt exhibits clear chemisorption. Substitutional boron doping substantially strengthens metal binding, but the enhancement is governed by local B motifs rather than the overall B concentration. Boron doping results in significantly stronger bonding compared to bonding to pristine graphene. Configurations in which multiple boron atoms surround an adsorption site deepen its electron-accepting character and promote charge-transfer assisted binding, especially for Mg, Zn and Cu. Pt, on the other hand, remains dominated by orbital hybridization, resulting in the absence of a strong linear correlation between adsorption energies and transferred charge. Mechanical strain in the range from −1 to +5% is found to have only a fine-tuning effect on metal adsorption, with energy changes below ~0.3 eV on B-doped surfaces. While strain does not fundamentally change the adsorption mechanism, it offers a secondary but potentially useful parameter for adjusting the performance and stability of metal atoms on graphene-based materials. Although strain does not substantially modify the near-Fermi electronic states of the substrate, in several cases it triggers adsorption-induced corrugation, particularly for strongly interacting metals such as Pt on B-rich compressed substrates. Oxidation introduces different types of interactions. Epoxy and hydroxyl groups strongly modify the binding, inducing MgO or MOH detachment from non-doped graphene. Boron doping stabilizes O-containing groups and hinders phase separation, except for Mg−OH which tends to separate regardless of doping. Oxidation is found to strengthen metal adsorption on B-graphene in most cases, relative to non-oxidized B-graphene. Among all the studied systems, only Mg adsorption on $C_{52}B_2O$ yields $|E_{ads}|$ exceeding the cohesive energy of the metal, suggesting a genuine thermodynamic preference for atomically dispersed Mg under high loading conditions. Finally, despite Pt adsorption never exceeding its cohesive energy, Pt anchored on B-doped and oxidized graphene shows substantial reactivity toward hydrogen, with $\Delta G(H_{ads})$ values between −0.23 and −0.58 eV, making these systems relevant for further exploration as model single-atom HER catalysts. Cu anchored on (oxidized) B-doped graphene appears not to be a good option due to strong CO binding, but on the other hand might be a promising candidate for CO oxidation. Overall, our results highlight that boron-doping motifs and oxygen functionalities are the primary levers for tuning metal-graphene interactions, while mechanical strain provides additional but comparatively minor modulation. These insights provide clear guidelines for the rational design of graphene-based support for metal-ion anchoring and single-atom catalysis.

## Acknowledgement

A.S.D. acknowledges the financial support provided by the Serbian Ministry of Science, Technological Development, and Innovations (contract no. 451-03-34/2026-03/200146), L'Oréal-UNESCO “For Women in Science” National Program in Serbia, and the Serbian Academy of Sciences and Arts (project F-49). The computations and data handling were enabled by resources provided by the National Academic Infrastructure for Supercomputing in Sweden (NAISS) at the National Supercomputer center (NSC) at Linköping University, partially funded by the Swedish Research Council through grant agreement No. NAISS 2024/5-718.

**SUPPLEMENTARY INFORMATION**

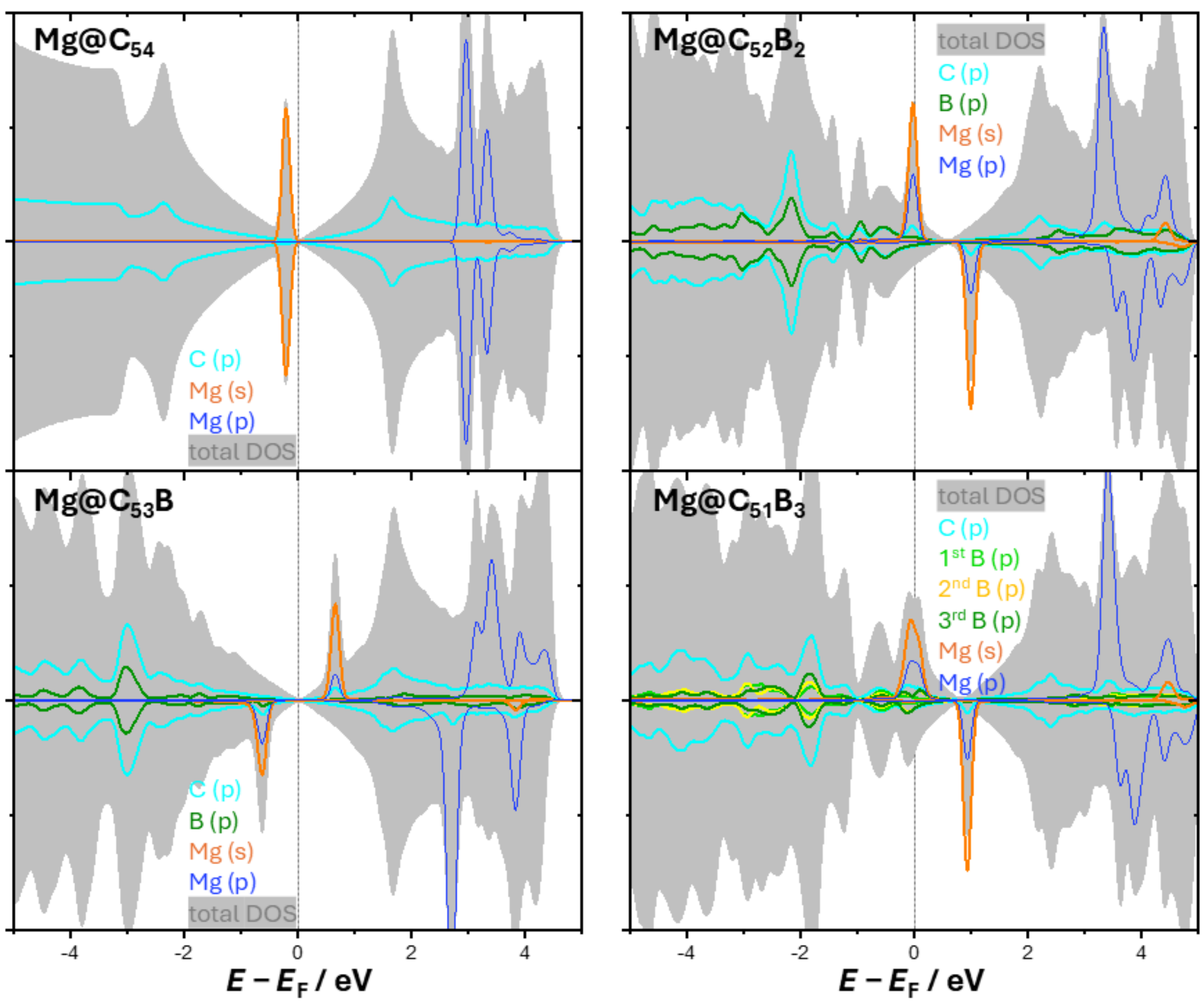

**Figure SI1.** Density of states (DOS) for Mg adsorbed on pristine graphene ($C_{54}$) and B-doped graphene ($C_{54-n}B_n$, $n \in \{0, 1, 2, 3\}$) at their preferred adsorption sites.

**Table SI1.** Metal adsorption energies onto $C_{54-n}B_n$, referred to isolated metal atoms ($E_{ads}$) and to metal atoms in pure metallic phase (**Δ***E*, calculated using metal cohesive energy). Metal aggregation is favored over metal-substrate interaction for $\Delta E > 0$ (numbers given in red).

| ***n*** | **Mg@$C_{54-n}B_n$** | | **Zn@$C_{54-n}B_n$** | | **Cu@$C_{54-n}B_n$** | | **Pt@$C_{54-n}B_n$** | |
|---|---|---|---|---|---|---|---|---|
| | **$E_{ads}$ / eV** | **Δ*E* / eV** | **$E_{ads}$ / eV** | **Δ*E* / eV** | **$E_{ads}$ / eV** | **Δ*E* / eV** | **$E_{ads}$ / eV** | **Δ*E* / eV** |
| **0** | −0.11 | 1.40 | −0.18 | 1.17 | −0.59 | 2.90 | −2.05 | 3.79 |
| **1** | −0.98 | 0.53 | −0.24 | 1.11 | −1.69 | 1.80 | −2.71 | 3.13 |
| **2** | −1.45 | 0.06 | −0.35 | 1.00 | −2.17 | 1.32 | −3.26 | 2.59 |
| **3** | −1.30 | 0.21 | −0.43 | 0.92 | −2.04 | 1.45 | −3.40 | 2.44 |

**Table SI2.** Metal adsorption energies onto epoxy-group containing boron-doped graphene, $C_{54-n}B_nO$, referred to isolated metal atoms ($E_{ads}$) and to metal atoms in pure metallic phase (**Δ***E*, calculated using metal cohesive energy). Metal aggregation is favored over metal-substrate interaction for $\Delta E > 0$ (numbers given in red).

| *n* | Mg@$C_{54-n}B_nO$ | | Zn@$C_{54-n}B_nO$ | | Cu@$C_{54-n}B_nO$ | | Pt@$C_{54-n}B_nO$ | |
|---|---|---|---|---|---|---|---|---|
| | $E_{ads}$ / eV | $\Delta E$ / eV | $E_{ads}$ / eV | $\Delta E$ / eV | $E_{ads}$ / eV | $\Delta E$ / eV | $E_{ads}$ / eV | $\Delta E$ / eV |
| **0** | −1.75 | −0.24 | −0.17 | 1.18 | −2.61 | 0.88 | −2.26 | 3.58 |
| **1** | −1.49 | 0.02 | −0.19 | 1.16 | −1.71 | 1.78 | −2.83 | 3.01 |
| **2** | −1.87 | −0.36 | −0.50 | 0.85 | −2.06 | 1.43 | −3.44 | 2.40 |
| **3** | −1.49 | 0.02 | −0.42 | 0.93 | −1.90 | 1.59 | −3.53 | 2.31 |

**Table SI3.** Metal adsorption energies onto hydroxyl-functionalized boron-doped graphene, $C_{54-n}B_nOH$, referred to isolated metal atoms ($E_{ads}$) and to metal atoms in pure metallic phase ($\Delta E$, calculated using metal cohesive energy). Metal aggregation is favored over metal-substrate interaction for $\Delta E > 0$ (numbers given in red).

| *n* | Mg@$C_{54-n}B_nOH$ | | Zn@$C_{54-n}B_nOH$ | | Cu@$C_{54-n}B_nOH$ | | Pt@$C_{54-n}B_nOH$ | |
|---|---|---|---|---|---|---|---|---|
| | $E_{ads}$ / eV | $\Delta E$ / eV | $E_{ads}$ / eV | $\Delta E$ / eV | $E_{ads}$ / eV | $\Delta E$ / eV | $E_{ads}$ / eV | $\Delta E$ / eV |
| **0** | −2.82* | −1.31* | −1.42* | −0.07* | −4.02* | −0.53* | −4.61* | 1.23* |
| **1** | −3.55 | −2.04 | −0.43 | 0.92 | −2.08 | 1.42 | −2.77 | 3.07 |
| **2** | −3.69 | −2.18 | −0.72 | 0.63 | −2.73 | 0.76 | −3.21 | 2.64 |
| **3** | −3.56 | −2.05 | −1.11 | 0.24 | −3.06 | 0.43 | −3.42 | 2.42 |

*MOH phase separation

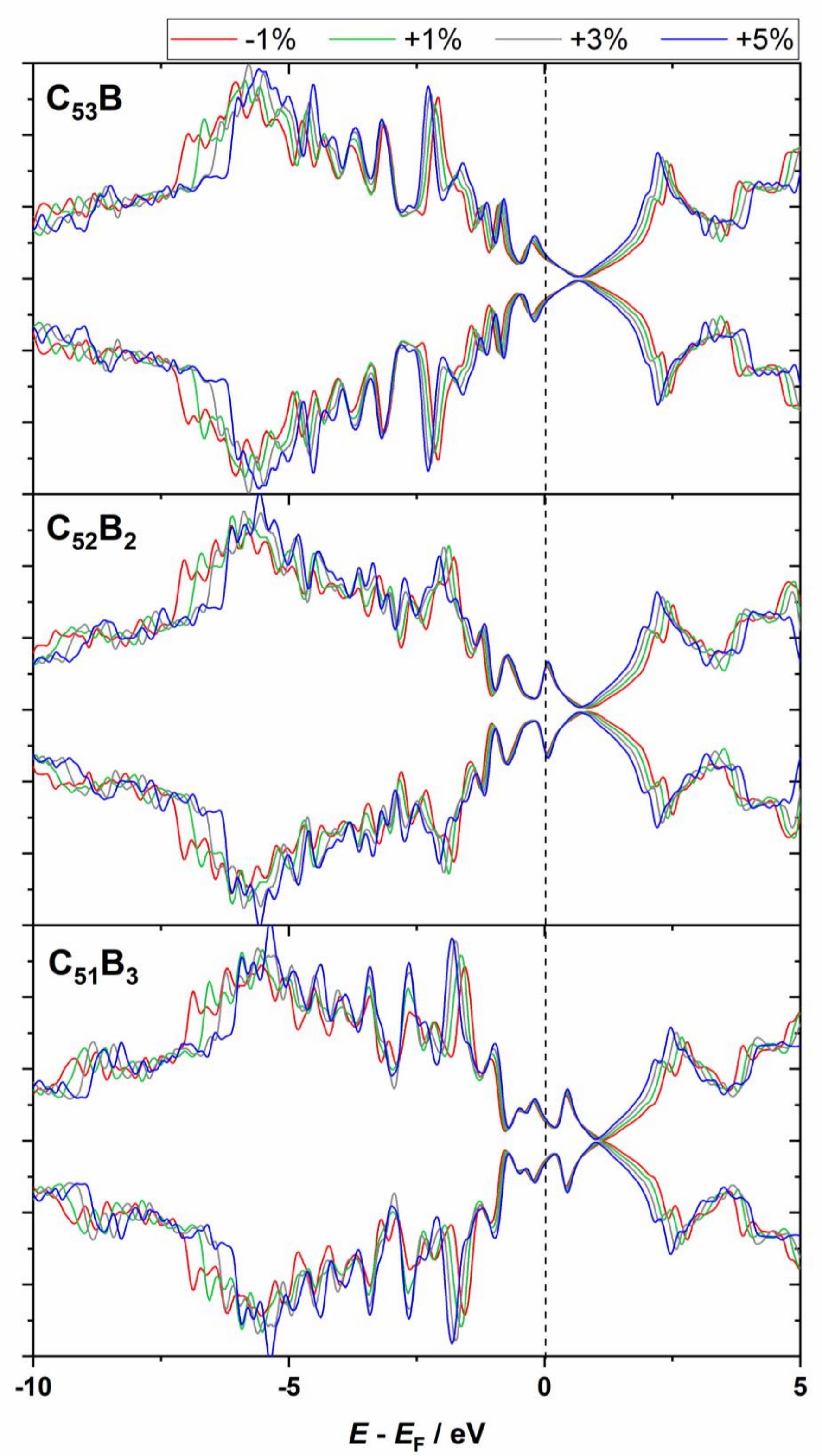

**Figure SI2.** Total densities of states of strained $C_{53}B$, $C_{52}B_2$ and $C_{51}B_3$ systems, with strain levels of −1%, +1%, +3% and +5%.

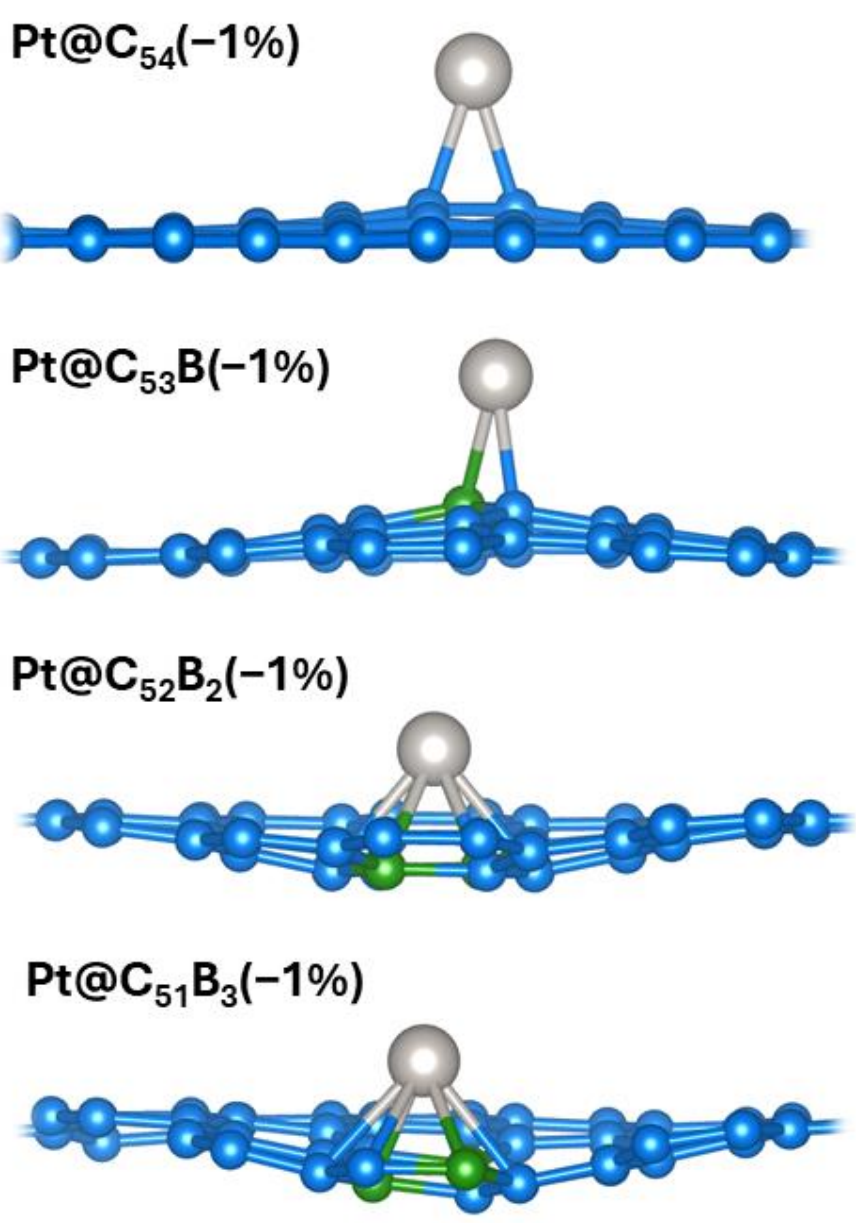

**Figure SI3.** Optimized structures of Pt interacting with $C_{54-n}B_n$ models compressed by 1% ($n \in \{0, 1, 2, 3\}$).

**Table SI4.** Adsorption of H onto Pt on $C_{54-n}B_n$ and their oxidized forms. Adsorption energies are calculated with respect to isolated H atom ($E_{ads}$(H)) and ½ of $H_2$ molecule ($E_{ads}(1/2H_2)$). The last column shows Gibbs free energy of H adsorption ($\Delta G(H_{ads})$).

| **System (H@Pt@subs)** | **$E_{ads}$(H) / eV** | **$E_{ads}(1/2H_2)$ / eV** | **$\Delta G(H_{ads})$ / eV** |
|---|---|---|---|
| H@Pt@$C_{54}$ | −2.82 | −0.57 | −0.33 |
| H@Pt@$C_{53}$B | −3.07 | −0.82 | −0.58 |
| H@Pt@$C_{52}B_2$ | −3.01 | −0.76 | −0.52 |
| H@Pt@$C_{51}B_3$ | −3.03 | −0.79 | −0.55 |
| H@Pt@$C_{54}$O | −2.71 | −0.47 | −0.23 |
| H@Pt@$C_{53}$BO | −2.92 | −0.67 | −0.43 |
| H@Pt@$C_{52}B_2$O | −2.89 | −0.65 | −0.41 |
| H@Pt@$C_{51}B_3$O | −2.80 | −0.55 | −0.31 |
| H@Pt@$C_{53}$BOH | −2.75 | −0.51 | −0.27 |
| H@Pt@$C_{52}B_2$OH | −2.83 | −0.58 | −0.34 |
| H@Pt@$C_{51}B_3$OH | −2.75 | −0.51 | −0.27 |

**Table SI5.** Adsorption energies of CO onto Cu on $C_{54-n}B_n$ and their oxidized forms, with corresponding bond lengths (*d*) between Cu and C from CO, and C and O in adsorbed CO.

| System (CO@Cu@subs) | $E_{ads}$(CO) / eV | *d*(Cu−C) / Å | *d*(C−O) / Å |
|---|---|---|---|
| CO@Cu@$C_{54}$ | −1.60 | 1.78 | 1.16 |
| CO@Cu@$C_{53}B$ | −1.99 | 1.78 | 1.15 |
| CO@Cu@$C_{52}B_2$ | −2.09 | 1.78 | 1.15 |
| CO@Cu@$C_{51}B_3$ | −2.04 | 1.78 | 1.15 |
| CO@Cu@$C_{53}BO$ | −2.08 | 1.77 | 1.15 |
| CO@Cu@$C_{52}B_2O$ | −1.90 | 1.78 | 1.15 |
| CO@Cu@$C_{51}B_3O$ | −1.94 | 1.78 | 1.15 |
| CO@Cu@$C_{53}BOH$ | −1.97 | 1.77 | 1.15 |
| CO@Cu@$C_{52}B_2OH$ | −1.79 | 1.78 | 1.15 |
| CO@Cu@$C_{51}B_3OH$ | −1.74 | 1.78 | 1.15 |

**Table SI6.** Adsorption energies of $CO_2$ onto Cu on $C_{54-n}B_n$ and their oxidized forms, with corresponding bond lengths (*d*) between Cu and C and/or O from $CO_2$, and C and O in adsorbed $CO_2$.

| System ($CO_2$@Cu@subs) | $E_{ads}$(CO) / eV | *d*(Cu-C/O) / Å | $d_1$(C−O) / Å | $d_2$(C−O) / Å | ∠(O−C−O) / ° |
|---|---|---|---|---|---|
| $CO_2$@Cu@$C_{54}$ | −0.37 | C 1.99<br>O 2.00 | 1.27 | 1.20 | 146.3 |
| $CO_2$@Cu@$C_{53}B$ | −0.49 | C 2.14<br>O 1.96 | 1.23 | 1.18 | 161.1 |
| $CO_2$@Cu@$C_{52}B_2$ | −0.50 | C 2.05<br>O 2.00 | 1.23 | 1.19 | 156.4 |
| $CO_2$@Cu@$C_{51}B_3$ | −0.49 | C 2.09<br>O 2.00 | 1.23 | 1.19 | 157.4 |
| $CO_2$@Cu@$C_{53}BO$ | −0.62 | C 2.00<br>O 1.97 | 1.24 | 1.19 | 154.3 |
| $CO_2$@Cu@$C_{52}B_2O$ | −0.55 | O 1.99 | 1.19 | 1.17 | 178.1 |
| $CO_2$@Cu@$C_{51}B_3O$ | −0.53 | O 1.99 | 1.19 | 1.17 | 179.0 |
| $CO_2$@Cu@$C_{53}BOH$ | −0.45 | O 2.01 | 1.18 | 1.17 | 179.4 |
| $CO_2$@Cu@$C_{52}B_2OH$ | −0.30 | O 2.12 | 1.19 | 1.17 | 179.6 |
| $CO_2$@Cu@$C_{51}B_3OH$ | −0.31 | O 2.08 | 1.18 | 1.17 | 179.6 |